\documentclass[journal,draftcls,onecolumn, 12pt]{IEEEtran}

\usepackage{psfrag}
\usepackage{amsmath}
\usepackage{amssymb}
\usepackage{graphicx}

\ifCLASSINFOpdf
\else
\fi
\usepackage{amssymb}
\usepackage{mdwlist}

\newtheorem{theo}{Theorem}

\begin{document}
%
\title{Performance and analysis of Quadratic Residue Codes of lengths less than 100}
%
%
%

\author{Yong~Li,~\IEEEmembership{Member,~IEEE,}
        Qianbin~Chen,~\IEEEmembership{Senior Member,~IEEE,}
        Hongqing~Liu,
        and~Trieu-Kien~Truong,~\IEEEmembership{Life Fellow,~IEEE}
\thanks{Yong~Li, Qianbin~Chen  and Hongqin~Liu are with the Key Lab
of Mobile Communication, Chongqing university of Posts and
Telecommunications, Chongqing,
 400065 China (e-mail: \{yongli, hongqingliu, chenqb\}@cqupt.edu.cn).}
 \thanks{Trieu-Kien~Truong is with the Department of Information Engineering, I-Shou
University, Kaohsiung Country 84001, Taiwan (e-mail:
truong@isu.edu.tw), and is also with the Department of Computer
Science and Engineering, National Sun Yat-sen University, Taiwan.}
}

\maketitle

\begin{abstract}
In this paper, the performance of quadratic residue (QR) codes of
lengths within 100 is given and analyzed when the hard decoding,
soft decoding, and linear programming decoding algorithms are
utilized.  We develop a simple method to estimate the soft
decoding performance, which avoids extensive simulations. Also, a
simulation-based algorithm is proposed to obtain the maximum
likelihood decoding performance of QR codes of lengths within 100.
Moreover, four important theorems are proposed to predict the
performance of the hard decoding and the maximum-likelihood
decoding in which they can explore some internal properties of QR
codes. It is shown that such four theorems can be applied to the
QR codes with lengths less than 100 for predicting the decoding
performance. In contrast, they can be straightforwardly
generalized to longer QR codes. The result is never seen in the
literature, to our knowledge. Simulation results show that the
estimated hard decoding performance is very accurate in the whole
signal-to-noise ratio (SNR) regimes, whereas the derived upper
bounds of the maximum likelihood decoding are only tight for
moderate to high SNR regions. For each of the considered QR codes,
the soft decoding is approximately 1.5 dB better than the hard
decoding. By using powerful redundant parity-check cuts, the
linear programming-based decoding algorithm, i.e., the ACG-ALP
decoding algorithm performs very well for any QR code. Sometimes,
it is even superior to the Chase-based soft decoding algorithm
significantly, and hence is only a few tenths of dB away from the
maximum likelihood decoding.
   \end{abstract}


%
\IEEEpeerreviewmaketitle

\section{Introduction}
%
%
%
%
\IEEEPARstart{T}{he} well-known binary quadratic residue (QR)
codes, first introduced by Prange \cite{Prange} in 1958, are a
nice family of cyclic binary BCH codes which have code rates
greater than or equal to 1/2 and generally have large minimum
distances so that most of the known QR codes are the best-known
codes. There are 11 binary QR codes in total with code lengths
less than 100; that is, 7, 17, 23, 31, 41, 47, 71, 73, 79, 89 and
97. So far the corresponding algebraic decoding algorithms for
these QR codes have been proposed by the authors given in [2-11].
Among them, the (89, 45, 17) QR code is the last decoded one.


In the past decades, the most widely used methods for decoding
binary QR codes are the Sylvester resultant \cite{Reed_2},
\cite{Reed_4} or Gr\"obner basis methods \cite{Chen_3}. These
methods can be utilized to solve the Newton identities that are
nonlinear and multivariate equations with high degree. However,
the calculations of identities require very high computational
complexity especially when the weight of the occurred error
pattern becomes large. Moreover, since different QR codes use
different sets of conditions to calculate the error locations, a
total enumeration of all conditions is impracticable for hardware
implementation. Therefore, these methods are only suitable for
relatively short QR codes such as codes with lengths 21, 31, and
41. Although the authors in \cite{Chen1994} developed an algebraic
decoding algorithm for decoding the (73, 37, 13) QR code based on
solving Newton identities, the simulation results were not
provided due to very complex computations. Later, the authors
\cite{Chen} have decoded QR codes using the well-known
Berlekamp-Massey (BM) algorithm. It is very efficient once the
needed consecutive syndromes are obtained. More generally, in
2001, He et al. \cite{He2001} developed an efficient matrix method
to determine the unknown syndromes by solving the equation
$\textrm{det}(S(I, J)) = 0$, where $\textrm{det}(\cdot)$ denotes
the determinant of a matrix and can be expressed as a polynomial
of some unknown syndrome. In particular, the (47, 24, 11) QR code
was successfully decoded by the BM algorithm, and the
corresponding algorithm was readily extended to decode the (71,
36, 11) , (79, 40, 15) and (97, 49, 15) QR codes \cite{Chang2003}.
Actually, such an algorithm can be utilized to decode any QR code
with an irreducible generator polynomial.

Among the foregoing 11 QR codes, there are two codes whose
generator polynomials are reducible, i.e., the (73, 37, 13) and
the (89, 45, 17) QR codes. In 2008, Truong et al. decoded the (89,
45, 17) QR code successfully by using a modification of the
inverse-free BM algorithm. Subsequently, Lin and Wang  et al.
\cite{Lin, Wang2013} further improved the decoding speed of the QR
code of length 89 by determining unknown syndromes quickly.
However, the conditions corresponding to error patterns with
different weights have still not been found. As a result, one
needs to try from $v = 1$ to $v = t$ errors before decoding
successfully or declaring a failure for any error pattern, where
$t$ represents the error-correcting capacity. As shown in
\cite{Wang2013}, the  linear programming decoding performance of
the (89, 45, 17) QR code is also investigated, and simulation
results show that the LP-based decoding performs slightly better
than the algebraic soft-decision decoding with comparable
complexity. Recently, an efficient algebraic decoding algorithm,
i.e., Lee et al.'s algorithm \cite{Lee2012} was proposed to decode
the (73, 37, 13) QR code up to six errors, which only needed to
calculate the unknown syndrome $S_5$ since it handled the
six-error case in terms of five-error case by inverting a bit of
the received vector. More recently, Li et al. \cite{Li2014}
proposed an improved algorithm for decoding the (73, 37, 13) QR
code. It was based on the hybrid unknown syndrome calculation
(HUSC) algorithm developed in \cite{Wang2013} and the modified
inverse-free BM algorithm \cite{Truong_2}. This further results in
a reduction of decoding complexity when compared with Lee et al.'s
algorithm.

The maximum likelihood (ML) decoding of a linear block code can be
described as an integer programming (IP) problem, which is
non-deterministic polynomial-time hard (NP-hard) \cite{Berlekamp}.
As an approximation to the ML decoding, the linear programming
(LP) decoding was first proposed by Feldman et al. \cite{Feldman}.
The number of constraints of the original LP decoding problem is
exponential in the maximum check node degree. Consequently, the
computational complexity may be prohibitively high even for some
small check degrees. To overcome this, an adaptive linear
programming (ALP) decoder \cite{Taghavi} was thus developed, which
reduced the number of constraints by adding only useful ones with
an adaptive and selective method. In addition, the performance of
the LP decoding can be further improved by adding more linear
constraints generated by redundant parity checks (RPC)
\cite{Taghavi, Tanatmis, Zhang, Miwa}. Although most of the LP
decoding algorithms were utilized to decode  LDPC codes, they
 also worked very well when used to
decode BCH codes, Golay codes, the (89, 45, 17), and the (73, 37,
13) QR codes, see \cite{Wang2013, Li2014, Tanatmis, Miwa}. One of
the interesting experimental results in \cite{Li2014} is that the
(73, 37, 13) QR code obtained a better performance than the (89,
45, 17) QR code with much fewer multiplications and additions when
the LP decoding was utilized despite the fact that the latter has
a larger minimum distance.

In this paper, first, we develop a simple algorithm to estimate
the SD decoding performance and a simulation-based method to
obtain the ML decoding performance for QR codes of lengths less
than or equal to 100. Next, we describe some new observations on
the performance of the hard-decision (HD) and the ML decoding for
QR codes and provide the corresponding proofs. Finally, We compare
the performance of the HD, soft-decision (SD), and LP decoding
algorithms for all the QR codes of lengths within 100 except the
(7, 4, 3) QR code (i.e., the classic Hamming code). The authors in
\cite{Draper2007} also proposed a mixed-integer LP method to
achieve the ML decoding of block codes with sparse parity-check
matrices based on the ALP decoding. However, such a method is not
suitable for QR codes since these codes have dense parity-check
matrices. Actually, the performance of the ALP decoding for QR
codes shown in section VI is very poor. Most components in the
solution are fractions when the ALP decoding terminates, which
results in that the required number of added integer constraints
needed in Draper et al.'s method \cite{Draper2007} is large and
thus difficult to solve.



The rest of this paper is organized as follows: The background of
the binary QR codes is introduced in Section II. Section III
reviews the algebraic decoding of QR codes based on solving Newton
identities and the inverse-free BM algorithm. In Section IV, the
LP decoding is briefly described.  Some new observations and the
corresponding proofs are provided in Section V. Simulation results
are presented in Section VI for the HD, SD, and LP decoding of the
QR codes. Finally, this paper concludes with a brief summary in
Section VII.


\section{Terminology and Background of the QR codes}

Let $n$ be a prime number of the form $n = 8l \pm 1$, where $l$ is
a positive integer. The set $Q_{n}$ of quadratic residues modulo
$n$ is the set of nonzero squares modulo $n$. That is,
\begin{equation}\label{eq:Qn}
Q_{n} = \{i| i \equiv  j^{2} \textrm{mod}\; n  \qquad \textrm{for}
 \;\; 1 \leq  j \leq n-1 \}.
\end{equation}
Let $m$ be the smallest positive integer such that $n$ divides
$2^{m} - 1$ and $\alpha$  is chosen a primitive element of the
finite field $GF(2^{m})$ such that each nonzero element of
$GF(2^{m})$ can be expressed as a power of  $\alpha$. Then the
element $\beta = \alpha^{u}$, where $u = (2^{m}-1)/n$, is a
primitive $n$th root of unity in $GF(2^{m})$. An $(n, k, d)$ QR
code which has the minimum distance $d$ is a cyclic code with the
generator polynomial $g(x)$ of the form  $g(x)=\prod_{i \in Q_{n}}
{(x-\beta^{i})}$.

For an $(n, k, d)$ QR code, an error pattern is said to be
correctable if its weight is less than or equal to the
error-correcting capacity  $t=\lfloor (d-1)/2 \rfloor$, where
$\lfloor x \rfloor$ denotes the greatest integer less than or
equal to $x$. Now, let the codeword $c(x) = c_{0} + c_{1}x +
\cdots + c_{n-1}x^{n-1}$ be transmitted through a noisy channel.
Also, let $ e(x) = e_{0} +e_{1}x + \cdots  + e_{n-1}x^{n-1} $ and
$r(x) = r_{0} + r_{1}x + \cdots + r_{n-1}x^{n-1}$ be the error
pattern occurred and the received vector, respectively. Then, the
received word has the form $r(x) = c(x) + e(x)$. The set of known
syndromes computed by evaluating $r(x)$ at the roots of $g(x)$ is
given by

\begin{equation}\label{eq:Si}
S_{i}=r(\beta^{i})=e(\beta^{i}), \qquad i \in Q_{n}.
\end{equation}

Assume that there are $v$ errors occurred in the received word
$r(x)$. Then, the error pattern has $v$ nonzero terms, namely
$e(x)=x^{l_{1}} + x^{l_{2}} + \cdots + x^{l_{v}}$, where $0\leq
l_{1}<l_{2}<\cdots<l_{v}\leq n-1$. The syndrome $S_i$ can be
written as $S_{i}=X_{1}^{i} + X_{2}^{i} + \cdots + X_{v}^{i}$,
where $X_{j}=\beta^{l_{j}}$ for $1\leq j \leq v$, are said to be
the error locators. If $i$ is not found in the set $Q_{n}$, the
syndrome $S_{i}$ is called an unknown syndrome.


Suppose that $v$ errors occur in the received word. The
error-locator polynomial is defined to be a polynomial of degree
$v$; that is,
$\sigma(x)=\prod_{j=1}^{v}(1+X_{j}x)=1+\sum_{j=1}^{v}\sigma_{j}x^{j}$
where $\sigma_{1} = X_{1} + \cdots + X_{v}$, $\sigma_{2} =
X_{1}X_{2} + \cdots + X_{v-1}X_{v}$, \ldots, and $\sigma_{v} =
X_{1}\cdots X_{v}$. One way to decode a QR code is to determine
the error-locator polynomial  $\sigma(x)$ and the Chien search is
then applied to find the roots of  $\sigma(x)$. The inverse-free
BM algorithm is known to be the most efficient method for
determining the error-locator polynomial. The error locations are
given by the inverse of the roots of  $\sigma(x)$ provided they
are no more than the error-correcting capacity $t$. In order to
use the inverse-free BM algorithm to decode a QR code up to $t$
errors, one needs to find, in sequence, the first 2$t$ consecutive
syndromes $S_{1}, S_{2}, \ldots , S_{2t}$. However, only the
syndromes whose indices are in $Q_n$ can be calculated directly
from $r(x)$; the others not determined directly from $r(x)$ are
unknown syndromes. Obviously, these known syndromes (resp.,
unknown syndromes) can be expressed as some powers of one or
several known syndromes (resp., unknown syndromes) that are the
so-called primary known syndromes (resp., unknown syndromes) of a
QR code.

Using a technique similar to that given in \cite{Chang2003}, a
strategy in \cite{Truong_2} was developed to obtain each of the
needed primary unknown syndromes. It is based on solving the roots
of the equation in a primary unknown syndrome with the
coefficients that can be expressed in terms of certain primary
known syndromes. Therefore, all of the unknown syndromes can be
calculated once the values of the primary unknown syndromes are
determined. The following is a brief review of the technique
mentioned in \cite{Chang2003}.

Assume that $v$ errors occur in the received word. Let $I$ =
\{$i_{1}$, $i_{2}$, $\ldots$ , $i_{v+1}$\} and $J$ = \{$j_{1}$,
$j_{2}$, $\ldots$ , $j_{v+1}$\} denote two subsets of \{0, 1, 2,
\ldots, $n$\}, respectively. These index subsets can be found by
an explicit use of the fast algorithm in \cite{Chen}. Next,
consider the matrix $S(I, J)$ of size ($v$+1)$\times$ ($v$+1),
given by

\begin{equation}\label{eq:Sij}
S(I, J) = \left[
\begin{array}{cccc}
S_{i_{1}+j_{1}} & S_{i_{1}+j_{2}} & \ldots & S_{i_{1}+j_{v+1}} \\
S_{i_{2}+j_{1}} & S_{i_{2}+j_{2}} & \ldots & S_{i_{2}+j_{v+1}} \\
\vdots & \vdots & \ddots & \vdots \\
S_{i_{v+1}+j_{1}} & S_{i_{v+1}+j_{2}} & \ldots &
S_{i_{v+1}+j_{v+1}}
\end{array} \right],
\end{equation}
where the summation of the indices of $S_{i}$'s is modulo $n$ and
the rank of $S(I, J)$ is at most $v$, which, in turn, implies

\begin{equation}\label{eq:detSij}
\textrm{det} \,S(I, J) = 0.
\end{equation}

If all of the unknown syndromes among the entries of $S(I, J)$
given in (\ref{eq:Sij}) can be expressed as some powers of one of
the primary unknown syndromes, say $S_{r}$, and if det($S(I, J)$)
is a nonzero polynomial in $S_{r}$, then the actual value of
$S_{r}$ is one of the roots of (\ref{eq:detSij}). In other words,
during the decoding process, one is able to calculate the value of
$S_{r}$ from the known syndromes.

\section{Algebraic decoding of QR codes}

Algebraic decoding of QR codes often includes three steps: 1)
Compute the unknown syndromes if needed. 2 ) Determine the
error-locator polynomial. 3) Apply the Chien search \cite{Chien}
to find the roots of the error-locator polynomial.
\subsection{Calculation of the error-locator polynomial based on solving Newton identities}

As mentioned in section II, one needs to compute the coefficients
\{$\sigma_j$\} of the error-locator polynomial $\sigma(x)$ where
$1 \leq j \leq t$. Here, $\{\sigma_j\}, 1 \leq j \leq t$ and
syndromes satisfy a series of Newton identities as follows:
\begin{eqnarray}
S_1 + \sigma_1 &=& 0 \nonumber\\
S_2 + S_1\sigma_1 &=& 0 \nonumber\\
S_3 + S_2\sigma_1 + S_1\sigma_2 + \sigma_3 &=& 0 \nonumber\\
S_4 + S_3\sigma_1 + S_2\sigma_2 + S_1\sigma_3 &=& 0 \nonumber\\
S_5+S_4\sigma_1+S_3\sigma_2+S_2\sigma_3+S_1\sigma_4+\sigma_5 &=& 0 \nonumber\\
S_6+S_5\sigma_1+S_4\sigma_2+S_3\sigma_3+S_2\sigma_4+S_1\sigma_5 &=& 0 \nonumber\\
S_7+S_6\sigma_1+S_5\sigma_2+S_4\sigma_3+S_3\sigma_4+S_2\sigma_5+S_1\sigma_6+\sigma_7 &=& 0 \nonumber\\
S_8+S_7\sigma_1+S_6\sigma_2+S_5\sigma_3+S_4\sigma_4+S_3\sigma_5+S_2\sigma_6+S_1\sigma_7 &=& 0 \nonumber\\
S_9+S_8\sigma_1+S_7\sigma_2+S_6\sigma_3+S_5\sigma_4+S_4\sigma_5+S_3\sigma_6+S_2\sigma_7+S_1\sigma_8+\sigma_9 &=& 0 \nonumber\\
S_{10}+S_9\sigma_1+S_8\sigma_2+S_7\sigma_3+S_6\sigma_4+S_5\sigma_5+S_4\sigma_6+S_3\sigma_7+S_2\sigma_8+S_1\sigma_9 &=& 0 \nonumber\\
& \vdots & \nonumber\\
S_{95}+S_{94}\sigma_1+S_{93}\sigma_2+S_{92}\sigma_3+S_{91}\sigma_4+S_{90}\sigma_5+S_{89}\sigma_6+S_{88}\sigma_7+S_{87}\sigma_8+S_{86}\sigma_9 &=& 0 \nonumber\\
S_{96}+S_{95}\sigma_1+S_{94}\sigma_2+S_{93}\sigma_3+S_{92}\sigma_4+S_{91}\sigma_5+S_{90}\sigma_6+S_{89}\sigma_7+S_{88}\sigma_8+S_{87}\sigma_9
&=& 0
\end{eqnarray}
Here, we only list the syndromes from $S_1$ to $S_{96}$ since the
longest one among QR codes of lengths within 100 is 97. There
exist the  known and unknown syndromes in the above Newton
identities such that one wishes to eliminate the unknown syndromes
when determining the coefficients $\{\sigma_j\}$. For example,
Elia's algorithm for decoding the (23, 12, 7) Golay code avoids
computing the unknown syndrome $S_5$ by utilizing the first four,
the seventh and the ninth identities \cite{Elia}. However, it is
very difficult to eliminate all the unknown syndromes in solving
Newton identities when code length is large, say $n \geq 47$.
Consequently, one needs to determine several or all unknown
syndromes before obtaining the coefficients $\{\sigma_j\}$ by
solving Newton identities.

\subsection{Computation of unknown syndromes}

Define a special sum of two subsets $I$ and $J$ to be a multi-set
as $I \oplus J = \{(i+j) \,\textrm{mod} \,n | i \in I, j \in J\}$,
where the star-sign ``$*$'' indicates that the set is a multi-set.
Its corresponding subsets $I$ and $J$ can be found by using a fast
algorithm in \cite{Chen}. After obtaining the subsets $I$ and $J$,
one can compute the unknown syndrome by solving the Eq.
(\ref{eq:detSij}).
 Sometimes, the degree of the resulting nonzero determinant polynomial of an unknown syndrome is so high
  that it is difficult to find the roots by using Chien's search. In this case, another determinant polynomial needs to be constructed.
  Euclid's method is then utilized to find the greatest common divisor (GCD) of
  two determinant polynomials $F(S_r)$ (with very low degree, e.g., degree is one) and thus one determines the unknown syndromes by solving the equation $F(S_r) = 0$.
 Furthermore, the subsets $I$ and $J$ provided by the algorithm in \cite{Chen} may generate a zero
 polynomial, i.e.,
$\textrm{det}(S(I, J)) \equiv 0$, thereby leading to a decoding
failure. For instance, when decoding the (73, 37, 13) QR code, the
subsets obtained from the algorithm in \cite{Chen} $I =
\{0,1,3,4,72\}$ and $J =\{0,2,72,71,70\}$ will result in zero
polynomials if $e(x) = 1 + x + x^2 + x^{68}, 1 + x + x^3 + x^{32}$
and $1 + x + x^{16} + x^{48}$, respectively. Towards this end,
additional subsets $I$ and $J$ are chosen so that nonzero
determinant
 polynomials can be  constructed. Based on solving Newton
 identities,  the decoding method   doesn't guarantee that the error patterns can
always be determined theoretically even if all the unknown
syndromes are obtained. Therefore all the possible error patterns
need to be verified before declaring whether the decoding
algorithm is correct or not. As a consequence,
 one needs to check $\sum_{i=1}^t C_i^{n}$ error
 patterns to validate the correctness of decoding algorithms.
 Obviously, it is impractical for those long QR codes. For
 example, $\sum_{i=1}^8 C_i^{89} = 78140695260$ error patterns in total needs to be checked for
 the (89, 45, 17) QR code. As shown in \cite{Li2014}, an improved algorithm
 was developed to search $I$ and $J$ which guarantees that the unknown
 syndromes can always be determined by solving Eq. (\ref{eq:detSij}). It is summarized in $Algorithm$ 1 as follows:

 $Algorithm$ 1:

\begin{enumerate}
 \item[Step 1:] Initially, Let $Q = Q_n \cup \{0\} \cup T_r$, where $T_r = \{t_r|t_r \equiv r \cdot 2^i \,\textrm{mod}\, n\}$.
  \item[Step 2:] Choose a subset $I = \{i_1, i_2, \ldots, i_{v+1}\} \subset Q$.
 \item[Step 3:] Check the number of elements in the intersection
 $(Q-i_1)\cap(Q-i_2)\cap \cdots \cap (Q-i_{v+1})$. If the
 cardinality of this intersection is less than $v+1$, return to
 Step 2.
  \item[Step 4:]  Choose a subset $J$ containing $v+1$ elements from the intersection in Step
  3. If all the possible sets $J$ have been chosen, return to Step 2.
  \item[Step 5:] If the intersection of the multi-set $I \oplus J$
  and $T_r$ is empty, return to Step 4.
  \item[Step 6:] Expand the polynomial $\textrm{det}(S(I, J))$ by using Laplace's method and check each monomial. If there exists
one monomial of the unknown syndrome whose coefficient is 1 and
whose power isn't different from that of other monomials, then
stop; otherwise, return Step 4.
\end{enumerate}

Since the degrees of determinant polynomials derived from
$Algorithm$ 1 may be high for large-weight error patterns, it is
too difficult to obtain the GCD of determinant polynomials with
very low degree. To overcome this, Gaussian elimination was
utilized to determine the unknown syndromes efficiently
\cite{Wang2013}. Its complexity is O($q\cdot (v+1)!$) for the
$v$-error case, where $q$ denotes the number of elements in the
finite field. Clearly, it may be still difficult or inefficient to
find the coefficients $\{\sigma_j\}$ by solving the Newton
identities for long QR codes even if all the unknown syndromes are
determined. For this reason, the inverse-free BM algorithm was
introduced to find the error-locator polynomials.

\subsection{Inverse-free BM decoding algorithm}

In fact, if the unknown syndromes are determined, the inverse-free
BM algorithm is more suitable for computing the error-locator
polynomial compared with solving Newton identities, especially
when $v$ is large. It only requires O($3t$) operations and is
summarized as follows:

\begin{enumerate}
 \item[Step 1:] Initially, Let $r^{(0)} = 1, C^{(0)}(x) = 1, A^{(0)}(x) = 1$, and $\ell^{(0)} = 1$.
  \item[Step 2:] Compute the discrepancy
  \begin{displaymath}
  \Delta^{(k)} = \sum_{j=1}^{\ell^{(k-1)}} c_{j-1}^{(k-1)}S_{k-j}+1.
  \end{displaymath}
 \item[Step 3:] Determine the error-locator polynomial
     \begin{displaymath}
     C^{(k)}(x) = r^{(k-1)}C^{(k-1)}(x) - \Delta^{(k)}A^{(k-1)}(x) \cdot x.
     \end{displaymath}
  \item[Step 4:]  Calculate the auxiliary variables
  \begin{eqnarray*}
  A^{(k)}(x) &=& \left \{ \begin{array}{ll}
x \cdot A^{(k-1)}(x),  & \textrm{if $\Delta^{(k)} = 0$ or if $2\ell^{(k-1)} > k-1$}\\
C^{(k-1)}(x), & \textrm{if $\Delta^{(k)} \neq 0$ and if $2\ell^{(k-1)} \leq k-1$ }.\\
\end{array} \right. \\
   \ell^{(k)} &=& \left \{ \begin{array}{ll}
\ell^{(k-1)},  & \textrm{if $\Delta^{(k)} = 0$ or if $2\ell^{(k-1)} > k-1$}\\
k-\ell^{(k-1)}, & \textrm{if $\Delta^{(k)} \neq 0$ and if $2\ell^{(k-1)} \leq k-1$ }.\\
\end{array} \right. \\
r^{(k)} &=& \left \{ \begin{array}{ll}
r^{(k-1)},  & \textrm{if $\Delta^{(k)} = 0$ or if $2\ell^{(k-1)} > k-1$}\\
\Delta^{(k)}, & \textrm{if $\Delta^{(k)} \neq 0$ and if $2\ell^{(k-1)} \leq k-1$ }.\\
\end{array} \right.
  \end{eqnarray*}
  \item[Step 5:] Update index $k = k + 1$ if $k \leq 2t$, then return to Step 2; otherwise, stop.
\end{enumerate}
Herein, the symbol $C^{(k)}(x)$ is the error-locator polynomial in
the stage $k$ and the syndromes $S_k, 1\leq k \leq 2t$, which can
be used to compute the discrepancy $\Delta^{(k)}$, are known. The
symbols $A^{(k)}(x), \ell^{(k)}$, and $r^{(k)}$ denote auxiliary
variables for determining the error-locator polynomial at the same
stage.

 It should be noted that $2t$
consecutive syndromes $S_1, \ldots, S_{2t}$ need to be calculated
in the inverse-free BM algorithm, whereas the needed syndromes may
not be successive when determining the error-locator polynomial by
solving Newton identities.

After obtaining the error-locator polynomial, the decoding can be
achieved straightforward by finding the roots of the error-locator
polynomial, which indicate the error locations. The inverse-free
BM-based algebraic decoding algorithm is suitable for any QR code
and its procedure can be seen in \cite{Chang2003, Truong_2,
Wang2013}. It is worth notice that the number of roots may not be
equal to the degree of the error-polynomial for QR codes with
reducible generator polynomials such as the (73, 37, 13) and the
(89, 45, 17) QR codes. In this case, error-locator polynomials are
correct when $v$ errors occur if and only if
$\textrm{deg}(\sigma(x))=v$ and the number of roots of $\sigma(x)
= 0$ equal to $v$ are satisfied simultaneously.

Combining the HD decoding and Chase-II algorithm, a complete soft
decoding of QR codes can be achieved. However, the Chase-II
decoding algorithm is much more time-consuming for a long QR code
because the HD decoder is repeated recursively for $2^{\lfloor d/2
\rfloor}$ times when decoding a received sequence. Towards this
end, a sufficient optimality condition \cite{Wang2013, Shu Lin}
was introduced to terminate the Chase decoding process more
rapidly. In order to further reduce the simulation time, a simple
method, called $Algorithm $ 2, is proposed to estimate the
performance of the SD decoding in this paper.

$Algorithm$ 2:

\begin{enumerate}
 \item[Step 1:] Initially, Let $\boldsymbol{c} = [c_0, c_1, \ldots,
 c_{n-1}]$ and $\boldsymbol{r} = [r_0, r_1, \ldots,
 r_{n-1}]$ denote a transmitted codeword and the corresponding
 received vector, respectively.
  \item[Step 2:] After BPSK modulation, the information bit becomes $(-1)^x$, where the input signal
 $x$ represents a binary 0 or 1 bit waveform for a communication system.
 Then the receiver decides the transmitted bit was 0 if $r_j > 0$
 or it decides the transmitted bit was 1 if $r_j < 0$. That is,
  \begin{displaymath}
  \hat{c}_j =  \left \{ \begin{array}{ll}
0, & \textrm{if $r_j > 0$ }\\
1, & \textrm{if $r_j < 0$.}\\
\end{array} \right.
  \end{displaymath}
 \item[Step 3:] Rearrange $|\boldsymbol{r}|$ in the reliability ascending order, where $|r_i|$ denotes the
 reliability of $i$-th element, and store the arranged indices in $\boldsymbol{\theta}$.
  \item[Step 4:]  Calculate the number of wrong bits in $\hat{\boldsymbol{c}}$, i.e., $\gamma_1$ and the maximum number
  of inverted wrong bits in the Chase decoding process, i.e., $\gamma_2$ as follows:
    \begin{enumerate}
   \item[]  $\gamma_1 = 0, \gamma_2 = 0$;
   \item[] $\textrm{for} \; i = 1: n$
   \item[] \hspace{0.4cm} $\textrm{if}\; (c_i \neq \hat{c}_i)$
   \item[] \hspace{0.8cm} $\gamma_1 = \gamma_1 + 1$;
   \item[]  \hspace{0.4cm} end
   \item[]  end
   \item[]  $\textrm{for} \; j = 1: \lfloor d/2 \rfloor$
   \item[]  \hspace{0.4cm}$\textrm{if}\; (c_{\theta[j]} \neq \hat{c}_{\theta[j]})$
   \item[]  \hspace{0.8cm} $\gamma_2 = \gamma_2 + 1$;
   \item[]  \hspace{0.4cm} end
   \item[]  end
  \end{enumerate}
    \item[Step 5:] If $\gamma_1 \leq \gamma_2 + t$, assume that
    the decoder decodes the received vector correctly; otherwise, declare a decoding failure.
\end{enumerate}
Actually, such an algorithm provides a lower bound of the
Chase-based SD decoding  algorithm by assuming that the
transmitted codeword can always be decoded once the new error
pattern after inverting one or a few bits of the received vector
is within the error correcting radius.

\section{Linear programming decoding}

As one of linear block codes, QR codes can also be represented by
the corresponding parity check matrices. There always exists a
direct relationship between the parity check matrix $H$ and the
so-called parity check polynomial $h(x)$. According to the
background of cyclic codes, $h(x)$ can be determined by solving
the following equation:
\begin{equation}
g(x)h(x) = x^n + 1.
\end{equation}

Since $h(x) = h_{k}x^k + \cdots + h_{1}x + h_0$, then we have

\begin{eqnarray*}
H = \left[
\begin{array}{ccccccccc}
h_0 & h_1 & \cdots &  & h_{k} & 0 & \cdots & & 0 \\
0 & h_0 & h_1 & \cdots &  & h_{k} & 0 & \cdots & 0 \\
\vdots & \vdots & \vdots & \vdots & \vdots & \vdots & \vdots & & \vdots \\
0 & \cdots & 0 & h_0 & h_1 & \cdots &  & & h_{k}
\end{array} \right]
\end{eqnarray*}

As is well known, the ML decoding process for any binary linear
code, denoted by an $m \times n$ check matrix $H$, can be written
as an optimization problem. That is,
\begin{equation}\label{eq:LP}
     \textrm{min} \; \boldsymbol{\gamma}^{T} \boldsymbol{u} \qquad
    \textrm{s.t.} \: \boldsymbol{u} \in  \textrm{conv($C$)}
\end{equation}
Here, $\boldsymbol{\gamma}$  is the cost vector obtained by the
log-likelihood ratios $\gamma_{j} = \log(P(y_{j}|u_{j}=0) /
P(y_{j}|u_{j}=1))$ for a given channel output $y_{j}$, $1 \leq j
\leq n$ and conv($C$) is the so-called codeword polytope.

As an approximation to the ML decoding process, Feldman et al.
\cite{Feldman} relaxed the codeword polytope onto the fundamental
polytope  so as to convert (\ref{eq:LP}) into a linear programming
problem. This polytope consists of both integral and nonintegral
vertices in which the former corresponds exactly to the codewords
of $C$. Thus, the LP relaxation yields the ML certificate
property; that is, if the LP decoder obtains an integral solution,
it is guaranteed to be an ML codeword.

In Feldman's original linear programming problem, the total number
of constraints, and hence the  computational complexity is
exponential in terms of the maximum check degree
$d_{i}^{\textrm{max}}, 1 \leq i \leq m$. This results in that the
explicit description of the fundamental polytope via parity
inequalities is inapplicable for high-density codes. To overcome
this, Feldman et al. proposed an equivalent formulation, which
requires O($N^{3}$) constraints \cite{Feldman}. Chertkov et al.
\cite{Chertkov} and Yang et al. \cite{Yang} then proposed an
alternative polytope, which has size linear in the code length and
the maximum check node degree.

  Taghavi and
Siegel \cite{Taghavi} introduced an adaptive approach to reduce
the computational complexity, called an adaptive linear
programming (ALP) decoding approach, which can be applicable to
high density codes instead of the direct implementation of the
original LP decoding algorithm. It should be noted that the ALP
decoder doesn't yield an improvement in terms of the frame error
rate (FER). In contrast, it has a very positive effect on the
decoding time because of converging with fewer constraints than
the original LP decoder.

Recently, Tanatmis et al. \cite{Tanatmis} proposed a separation
algorithm (SA) in order to improve the error-correcting
performance of the LP decoding, which is abbreviated as the SALP
algorithm henceforward. More recently, Zhang and Siegel
\cite{Zhang} developed a novel decoder that combines the new
adaptive cut-generating (ACG) algorithm with the ALP algorithm,
called ACG-ALP decoder. Here the details of the SALP and the
ACG-ALP decoders are omitted due to limited space.

Based on the ACG-ALP decoding, a simulation-based method to obtain
the ML decoding performance for QR codes of lengths within 100 is
developed. It can be summarized by the following steps:

\begin{enumerate}
 \item[1)] Initially, decode the received vector by utilizing the ACG-ALP
 decoder.
  \item[2)]  If an ML codeword is output, then stop;
  otherwise, use all the constraints of the ACG-ALP
 decoder as the cut cool to construct an integral programming
 decoder and solve it.
\end{enumerate}

Such an idea is based on a conjecture that the solution of the
ACG-ALP decoder is not far away from the ML codeword when a
decoding failure is declared. Our simulations verify this
conjecture and it can be much more time-consuming if the ACG-ALP
decoder is replaced by other LP-based ones, say the SALP and the
ALP decoders.
\section{Some new observations}

Based on analyzing  the error-rate performance of the HD and ML
decoding of QR codes, the  following theorems are obtained.

\begin{theo}\label{prop:PEHD}
Let an $(n, k, d)$ QR code be transmitted over AWGN channels, the
frame error rate (sometimes called the codeword error rate) of the
HD decoding can be computed by
\begin{equation}\label{eq:pE_HD}
p_E = 1 - [(1-\bar{p})^{n} + C_{n}^1 \cdot \bar{p} \cdot
(1-\bar{p})^{n-1}+ \cdots +  C_{n}^{t} \bar{p}^{t} \cdot
(1-\bar{p})^{n-t}],
\end{equation}
where $\bar{p}$ indicates the average bit error probability.
\end{theo}

\begin{IEEEproof}
Without loss of generality,  assume that all-zero codewords are
transmitted. Then the channel model can be written as
\begin{displaymath}
y_k = 1 + n_k, \; k = 1, \ldots, n,
\end{displaymath}
where $y_k$ represents the channel received value and  $n_k$ is a
Gaussian random variable with zero mean and $\sigma^2$ variance.

Given $y_k$, the bit error probability of the $k$-th coded bit
$p_k$, see \cite{Lin2009}, is given by the following formula:
\begin{displaymath}
p_k = \frac{1}{1+e^{2|y_k| / \sigma^2}}, -\infty < y_k < +\infty.
\end{displaymath}

Since $p_k$ is a function of the received information $y_k$ which
is a discrete-time conditional Gaussian random sequence, the mean
of the error probability is given as follows:

\begin{eqnarray}
\bar{p} & = & E(p_k) = \int_{-\infty}^{+\infty} p_k f(y_k)d y_k \nonumber\\
&=& \frac{1}{\sqrt{2\pi}\sigma} \int_{-\infty}^{+\infty}
\frac{1}{1+e^{2|x|/ \sigma^2}} \cdot
 e^{-\frac{(x - 1)^2}{2\sigma^2}} d x
\end{eqnarray}
The explicit value of such an integration cannot be obtained,
whereas its numerical solution can be calculated by Simpson
quadrature. Obviously, it is a monotonically increasing function
on $\sigma^2$.

For the $(n, k, d)$ QR code, the probability that the HD decoding
makes the correct decision is given by
\begin{equation}\label{eq:ps_HD}
p_s = (1-\bar{p})^{n} + C_{n}^1 \cdot \bar{p} \cdot
(1-\bar{p})^{n-1}+ \cdots +  C_{n}^{t} \bar{p}^{t} \cdot
(1-\bar{p})^{n-t}
\end{equation}

$p_E$ in (\ref{eq:pE_HD}) immediately follows from Eq.
(\ref{eq:ps_HD}). This complete the proof of Theorem 1.

\end{IEEEproof}

\begin{theo}\label{prop:HD}
Consider the $(n_1, k_1, d_1)$ and the $(n_2, k_2, d_2)$ two QR
codes of lengths within 100  transmitted over AWGN channels. If
$d_1 = d_2$, then the QR code with the shorter length has a better
performance under the HD decoding.
\end{theo}

\begin{figure}[htbp]
\centerline{
\includegraphics[width=4.0in]{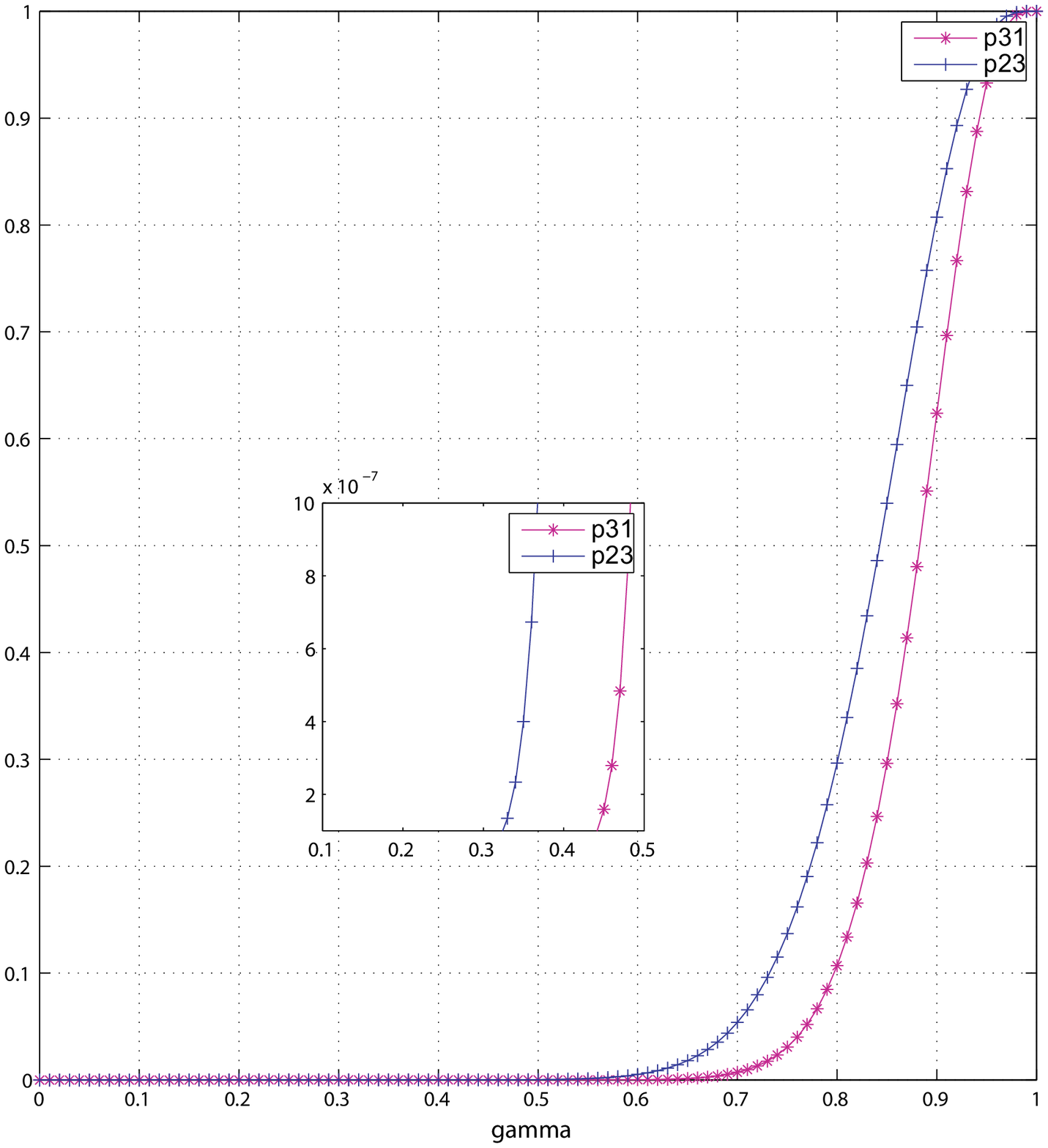}}
  \caption{The curves of $p_{s, 23}$ (p23 in the legend) and $p_{s, 31}$ (p31 in the legend).}
  \label{fig:theorem}
\end{figure}

\begin{IEEEproof}
 There are three pairs of QR
codes with the same correcting capacities; that is, the (23, 12,
7) and the (31, 16, 7) QR codes, the (47, 24, 11) and the (71, 36,
11) QR codes, the (79, 40, 15) and the (97, 49, 15) QR codes. In
the following, we first prove  the case of the (23, 12, 7) and the
(31, 16, 7) QR codes as an example.

Let $p_{s,23}$ and $p_{s,31}$ be the successful probabilities of
HD decoding of the (23, 12, 7) and  the (31, 16, 7) QR codes
respectively. Then according to Eq. (\ref{eq:ps_HD}), we have
\begin{eqnarray*}
p_{s,23} &=& (1-\bar{p})^{23} + C_{23}^1 \cdot \bar{p} \cdot
(1-\bar{p})^{22} + C_{23}^2 \cdot \bar{p}^2 \cdot (1-\bar{p})^{21}
+ C_{23}^3 \cdot \bar{p}^3 \cdot (1-\bar{p})^{20}, \\
p_{s,31} &=& (1-\bar{p})^{31} + C_{31}^1 \cdot \bar{p} \cdot
(1-\bar{p})^{30} + C_{31}^2 \cdot \bar{p}^2 \cdot (1-\bar{p})^{29}
+ C_{31}^3 \cdot \bar{p}^3 \cdot (1-\bar{p})^{28}
\end{eqnarray*}

Let $\gamma = 1-\bar{p}$. Then $p_{s,23}$ and $p_{s,31}$ are two
functions of $\gamma$ as shown in Fig. \ref{fig:theorem}. Both of
which have only two intersection points on the closed interval [0,
1], i.e., $\gamma = 0$ and $\gamma = 1.0$, and the former is
always larger than the latter on the open interval (0, 1). Hence,
the (23, 12, 7) QR code performs better than the (31, 16, 7) QR
code under the HD decoding.

In a similar manner as the proof of the first pair of QR codes,
another two pairs of QR codes are thus proved.
\end{IEEEproof}

\begin{theo}\label{prop:PE_MLD}
Consider an $(n, k, d)$ QR code with a weight numerator $A(z) =
\sum_i \lambda_i z^{\omega_i}$. An  upper bound of the FER of the
ML decoding can be expressed as
\begin{equation} \label{prop:FERMLD}
p_{E}(ML) \leq 1-  \prod_{i, \omega_i \neq 0} (\frac{1}{2}
\textrm{erfc}(\frac{-\sqrt{\omega_i}}{\sqrt{2}
\sigma}))^{\lambda_i}
\end{equation}
\end{theo}
Here, the `=' is approximately valid  and thus the FER performance
of the ML decoding can be estimated by the right hand side of
(\ref{prop:FERMLD}) in high SNR regimes.

\begin{IEEEproof}
Now and in the sequel, we use the same symbol definitions of
$Theorem$ 1 and assume that all-zero codewords are always
transmitted. According to Eq. (\ref{eq:LP}), the probability that
the ML decoding makes the correct decision is given by
\begin{eqnarray} \label{prop:PMLD}
p_{ML} &=& \textrm{Pr}\{\forall \; \boldsymbol{u} \in \cal{C} \;
\textrm{satisfies} \; \boldsymbol{y}^T \cdot \boldsymbol{u} >
\textrm{0}, \boldsymbol{u} \neq
\boldsymbol{0}\} \nonumber\\
&\geq &  \prod_{i, \omega_i \neq 0} \textrm{Pr}\{\boldsymbol{y}^T
\cdot \boldsymbol{u}
>
\textrm{0}, \textrm{the weight of $\boldsymbol{u}$ is $\omega_i$}\} \nonumber\\
&=& \prod_{i, \omega_i \neq 0} (\textrm{Pr}(\boldsymbol{y}^T \cdot
\boldsymbol{u}_i >
\textrm{0}))^{\lambda_i} \nonumber\\
&=& \prod_{i, \omega_i \neq 0} (\textrm{Pr}(\sum_{j=1}^{\omega_i}
y_j  > 0))^{\lambda_i},
\end{eqnarray}
where $\lambda_i$ indicates the number of codewords with weight
$\omega_i$, and $\boldsymbol{u}_i$  is a codeword whose weight is
$\omega_i$. The $`\geq'$ in (\ref{prop:PMLD}) results from the
formula $\textrm{Pr}(A,B) = \textrm{Pr}(A)\cdot \textrm{Pr}(B|A)
\geq \textrm{Pr}(A)\cdot \textrm{Pr}(B)$ where $A$ and $B$ denote
two events, respectively, and the first $`='$  is valid because
the weight of $\boldsymbol{u}_i$ is $\omega_i$.

In general, assume that $\{y_j, j = 1, \ldots, n\}$ is an i.i.d
sequence. Then $\sum_{j=1}^{\omega_i} y_j$ satisfies a Gaussian
distribution with mean = $\omega_i$ and variance = $\omega_i
\sigma^2$, where $\sigma^2$ is the variance of a noise variable in
AWGN channel. As a result, we have
\begin{equation} \label{prop:temp}
\textrm{Pr}(\sum_{j=1}^{\omega_i} y_j > 0) = \int_{0}^{+\infty}
\frac{1}{\sqrt{2\pi \omega_i}\sigma} e^{-\frac{(x -
\omega_i)^2}{2\omega_i \sigma^2}} dx = \frac{1}{2}
\textrm{erfc}(\frac{-\sqrt{\omega_i}}{\sqrt{2} \sigma}),
\end{equation}
where erfc($x$) is the error discrepancy function. A substitution
of Eq. (\ref{prop:temp}) into (\ref{prop:PMLD}) yields

\begin{equation} \label{prop:MLDC}
p_{ML} \geq \prod_{i, \omega_i \neq 0} (\frac{1}{2}
\textrm{erfc}(\frac{-\sqrt{\omega_i}}{\sqrt{2}
\sigma}))^{\lambda_i}
\end{equation}
Accordingly, $p_{E}(ML)$ given in (\ref{prop:FERMLD}) can be
directly deduced from (\ref{prop:MLDC}).

\end{IEEEproof}

In fact, Eq. (\ref{prop:FERMLD}) is very useful and efficient when
predicting the performance of the ML decoding. As shown in the
next section, it is very tight in high SNR regimes and can thus be
utilized to calculate the ML decoding performance of any block
code once the corresponding weight numerator is known.

\begin{theo}\label{prop:MLD}
Consider the $(n_1, k_1, d_1)$ and the $(n_2, k_2, d_2)$ QR codes
of lengths within 100  transmitted over AWGN channels. If $d_1 =
d_2$, then the QR code with the longer length has a better
performance under the ML decoding in high SNR regimes.
\end{theo}

\begin{IEEEproof}
Again, we first use the (23, 12, 7) and the (31, 16, 7) QR codes
as example to complete the proof. According to the results in
\cite{Truong2005}, the weight numerators of two QR codes of
lengths 23 and 31, namely $A_{23}(z)$ and $A_{31}(z)$, can be
written as follows:
\begin{eqnarray*}
A_{23}(z) &=& 1 + 253z^7 + 506z^8 + 1288z^{11} + 1288z^{12} +
506z^{15} +
253z^{16} + z^{23}. \\
A_{31}(z) &=& 1 + 155z^7 + 465z^8 + 5208z^{11} + 8680z^{12} +
18259z^{15} + 18259z^{16} \\
& & + 8680z^{19} + 5208z^{20} + 465z^{23} + 155z^{24} + z^{31}.
\end{eqnarray*}
With the aid of Eq. (\ref{prop:MLDC}), the  probability that the
ML decoder makes the correct decision for the (23, 12, 7) and the
(31, 16, 7) QR codes can be obtained by
\begin{eqnarray*}
p_{23-ML} &=&
(\frac{1}{2}\textrm{erfc}(\frac{-\sqrt{7}}{\sqrt{2}\sigma_1}))^{253}
 \cdot
(\frac{1}{2}\textrm{erfc}(\frac{-\sqrt{8}}{\sqrt{2}\sigma_1}))^{506}
\cdot
(\frac{1}{2}\textrm{erfc}(\frac{-\sqrt{11}}{\sqrt{2}\sigma_1}))^{1288}
\cdot
(\frac{1}{2}\textrm{erfc}(\frac{-\sqrt{12}}{\sqrt{2}\sigma_1}))^{1288}\\
& & \cdot
(\frac{1}{2}\textrm{erfc}(\frac{-\sqrt{15}}{\sqrt{2}\sigma_1}))^{506}
\cdot
(\frac{1}{2}\textrm{erfc}(\frac{-\sqrt{16}}{\sqrt{2}\sigma_1}))^{253}
\cdot
(\frac{1}{2}\textrm{erfc}(\frac{-\sqrt{23}}{\sqrt{2}\sigma_1}))) \\
p_{31-ML} & = &
(\frac{1}{2}\textrm{erfc}(\frac{-\sqrt{7}}{\sqrt{2}\sigma_2}))^{155}
\cdot
(\frac{1}{2}\textrm{erfc}(\frac{-\sqrt{8}}{\sqrt{2}\sigma_2}))^{465}
\cdot
(\frac{1}{2}\textrm{erfc}(\frac{-\sqrt{11}}{\sqrt{2}\sigma_2}))^{5208}
\cdot
(\frac{1}{2}\textrm{erfc}(\frac{-\sqrt{12}}{\sqrt{2}\sigma_2}))^{8680}\\
& & \cdot
(\frac{1}{2}\textrm{erfc}(\frac{-\sqrt{15}}{\sqrt{2}\sigma_2}))^{18259}
\cdot
(\frac{1}{2}\textrm{erfc}(\frac{-\sqrt{16}}{\sqrt{2}\sigma_2}))^{18259}
\cdot
(\frac{1}{2}\textrm{erfc}(\frac{-\sqrt{19}}{\sqrt{2}\sigma_2})))^{8680}\\
& &\cdot
(\frac{1}{2}\textrm{erfc}(\frac{-\sqrt{20}}{\sqrt{2}\sigma_2})))^{5208}
 \cdot
(\frac{1}{2}\textrm{erfc}(\frac{-\sqrt{23}}{\sqrt{2}\sigma_2})))^{465}
\cdot
(\frac{1}{2}\textrm{erfc}(\frac{-\sqrt{24}}{\sqrt{2}\sigma_2})))^{155}
\cdot
(\frac{1}{2}\textrm{erfc}(\frac{-\sqrt{31}}{\sqrt{2}\sigma_2}))).
\end{eqnarray*}
Here, $\sigma_1$ and $\sigma_2$ represent the channel standard
deviation when the (23, 12, 7) and the (31, 16, 7) QR codes are
transmitted over a channel that is corrupted by additive noise.
 In the high SNR region, each term of $p_{23-ML}$ and $p_{31-ML}$
approaches 1 and the value of the term with a larger weight is
also often larger. Thus, the values of both of which are mainly
dominated by first two monomials, i.e.,
\begin{eqnarray*}
p_{23-ML} & \approx &
(\frac{1}{2}\textrm{erfc}(\frac{-\sqrt{7}}{\sqrt{2}\sigma_1}))^{253}
\cdot
(\frac{1}{2}\textrm{erfc}(\frac{-\sqrt{8}}{\sqrt{2}\sigma_1}))^{506}\\
&=&
(1-\frac{1}{2}\textrm{erfc}(\frac{\sqrt{7}}{\sqrt{2}\sigma_1}))^{253}
\cdot
(1-\frac{1}{2}\textrm{erfc}(\frac{\sqrt{8}}{\sqrt{2}\sigma_1}))^{506}\\
& \approx &
(1-\frac{253}{2}\textrm{erfc}(\frac{\sqrt{7}}{\sqrt{2}\sigma_1}))
\cdot
(1-\frac{506}{2}\textrm{erfc}(\frac{\sqrt{8}}{\sqrt{2}\sigma_1})).\\
 p_{31-ML} & \approx &
(\frac{1}{2}\textrm{erfc}(\frac{-\sqrt{7}}{\sqrt{2}\sigma_2}))^{155}
\cdot
(\frac{1}{2}\textrm{erfc}(\frac{-\sqrt{8}}{\sqrt{2}\sigma_2}))^{465}\\
&=&
(1-\frac{1}{2}\textrm{erfc}(\frac{\sqrt{7}}{\sqrt{2}\sigma_2}))^{155}
\cdot
(1-\frac{1}{2}\textrm{erfc}(\frac{\sqrt{8}}{\sqrt{2}\sigma_2}))^{465}\\
& \approx &
(1-\frac{155}{2}\textrm{erfc}(\frac{\sqrt{7}}{\sqrt{2}\sigma_2}))
\cdot
(1-\frac{465}{2}\textrm{erfc}(\frac{\sqrt{8}}{\sqrt{2}\sigma_2})).
\end{eqnarray*}
Here, the second $`\approx'$ for $p_{23-ML} (p_{31-ML})$ is valid
according to the Taylor approximation.

 Now, let $R_1 = k_1/n_1$  and $R_2 = k_2/n_2$ denote the rates of the
 $(n_1, k_1, d_1)$ and the $(n_2, k_2, d_2)$ QR codes, respectively. If $R_1 = R_2 \approx \frac{1}{2}$, then
 $\sigma_1 \approx \sigma_2 = \sigma$ at the same SNR level. As a
consequence, we have
\begin{eqnarray*}
p_{23-ML} - p_{31-ML} &\approx&
(1-\frac{253}{2}\textrm{erfc}(\frac{\sqrt{7}}{\sqrt{2}\sigma}))
\cdot
(1-\frac{506}{2}\textrm{erfc}(\frac{\sqrt{8}}{\sqrt{2}\sigma}))
\\
& &-
(1-\frac{155}{2}\textrm{erfc}(\frac{\sqrt{7}}{\sqrt{2}\sigma}))
\cdot
(1-\frac{465}{2}\textrm{erfc}(\frac{\sqrt{8}}{\sqrt{2}\sigma})) \\
& < &
(1-\frac{253}{2}\textrm{erfc}(\frac{\sqrt{7}}{\sqrt{2}\sigma}))
\cdot
(1-\frac{465}{2}\textrm{erfc}(\frac{\sqrt{8}}{\sqrt{2}\sigma})) \\
& & -
(1-\frac{155}{2}\textrm{erfc}(\frac{\sqrt{7}}{\sqrt{2}\sigma}))
\cdot
(1-\frac{465}{2}\textrm{erfc}(\frac{\sqrt{8}}{\sqrt{2}\sigma}))\\
& = & (-49 \textrm{erfc}(\frac{\sqrt{7}}{\sqrt{2}\sigma})) \cdot
(1-\frac{465}{2}\textrm{erfc}(\frac{\sqrt{8}}{\sqrt{2}\sigma})) <
0.
\end{eqnarray*}

We must highlight that the proof is not very strict because $R_2$
is slightly higher than $R_1$. However, it is more rigorous for
the $(n_1+1, k_1, d_1+1)$ and the $(n_2+1, k_2, d_2+1)$ extended
QR codes because both rates $R_1$ and $R_2$ are equal to 0.5.
\end{IEEEproof}

The above four theorems will be helpful in the following
simulations.

\section{Simulation results}

Simulations of different decoding algorithms for QR codes in AWGN
with BPSK modulation are conducted using the C++ programming
language. The Simplex method is used as our LP solver and CPLEX
11.0 \cite{CPLEX} is utilized as an IP solver. In this paper, only
10 QR codes except the Hamming code are considered since it is a
well-known code and is used extensively in communication systems.
A frame error rate (FER) is declared for each simulation when
collecting 100 erroneous codewords including incorrect codewords
and pseudo-codewords.  The simulation-based method described in
section IV is utilized to obtain the performance curve of the ML
decoding. It should be noted that only a lower bound of the SD
decoding obtained from $Algorithm$ 2 is given for each QR code
whose length belongs to the set $\{17, 31, 41, 47, 71, 79, 97\}$.
%

\begin{figure}[htbp]
\centerline{
\includegraphics[width=4.0in]{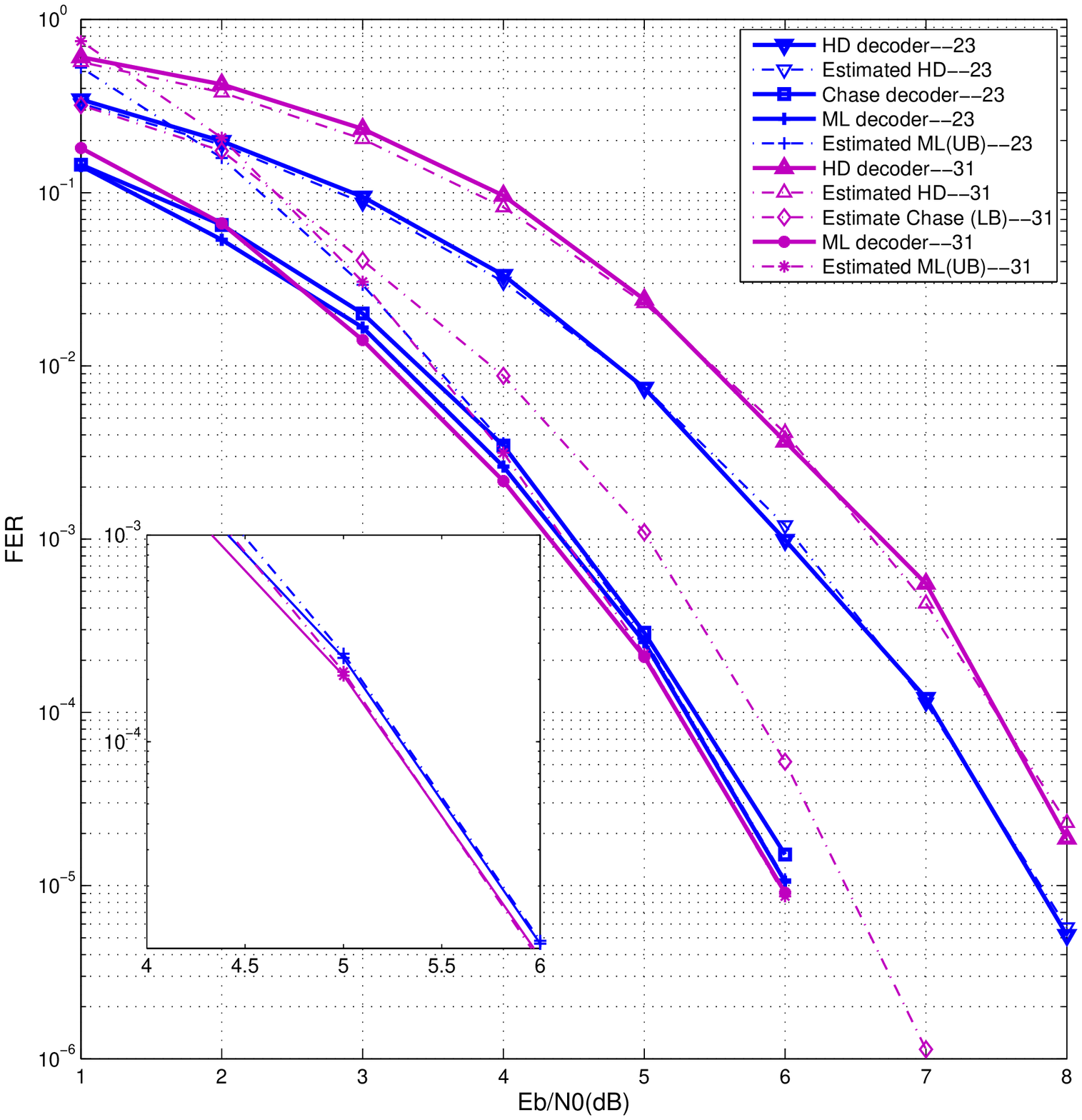}}
  \caption{FER performance comparison of the (23, 12, 7) and the (31, 16, 7) QR codes in AWGN channel. Estimated Chase (LB): the lower bound of the Chase decoding
  obtained from Algorithm 2; Estimated ML (UB): the upper bound of the ML decoding obtained from Theorem 3. The integers in the legend denote the code lengths of QR codes.}
  \label{fig:23-31}
\end{figure}

In comparison to the frame error rate  (FER) performance of the
(23, 12, 7) QR code with the (31, 16, 7) QR code when different
decoding algorithms are used, as illustrated in Fig.
\ref{fig:23-31}, the curves of the decoding of the both QR codes
by using the Chase-II based SD decoding algorithm \cite{Chase} are
approximate 1.5 dB away from that by using the HD decoding
algorithm at a FER of $10^{-5}$. The SD decoding performance for
the (23, 12, 7) code is obtained by simulations, whereas a lower
bound of SD decoding of the (31, 16, 7) QR code is provided
according to the method described in section III. One observes
from the figure that the HD decoding of the (23, 12, 7) QR code
performs better than that of the (31, 16, 7) QR code, which is
consistent with $Theorem$ \ref{prop:HD}. It is also the case for
the SD decoding since the kernel of the Chase-II based SD decoding
is the HD decoding. As presented in $Theorem$ \ref{prop:MLD}, the
latter is slightly superior to the former under the ML decoding
since the former has more codewords with lower weights. Both the
HD and the ML decoding performance curves of both codes calculated
by $Theorem$s \ref{prop:PEHD} and \ref{prop:PE_MLD} are also
plotted in Fig. \ref{fig:23-31}. It can also be readily seen that
$Theorem$ \ref{prop:PE_MLD} matches the simulation results very
well for moderate to high SNR regimes, whereas $Theorem$
\ref{prop:PEHD} is nearly in agreement with the simulation in the
whole simulated SNR regions.

\begin{figure}[htbp]
\centerline{
\includegraphics[width=4.0in]{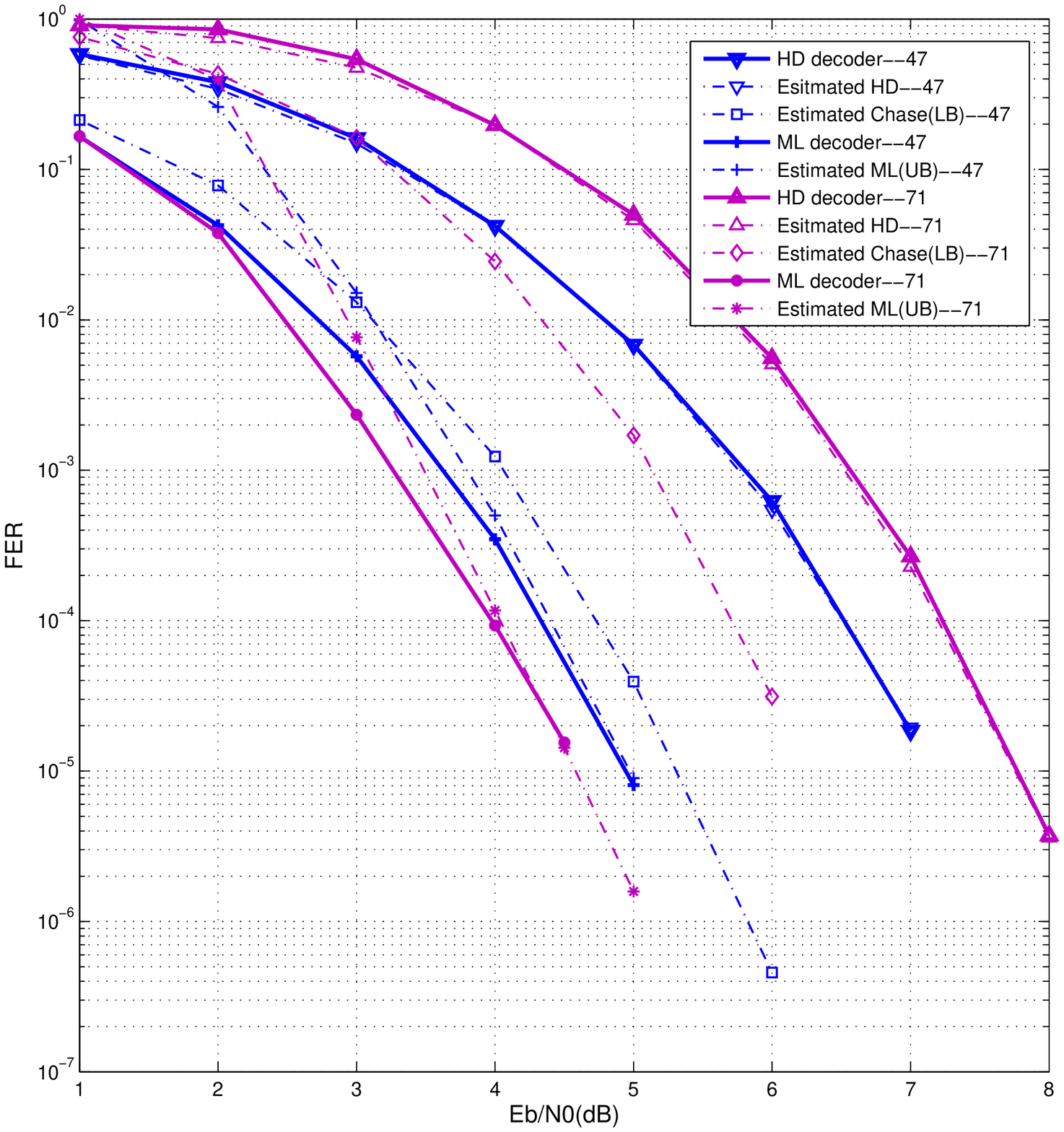}}
  \caption{FER performance comparison of the (47, 24, 11) and the (71, 36, 11) QR codes in AWGN channel. Estimated Chase (LB): the lower bound of the Chase decoding
  obtained from Algorithm 2; Estimated ML (UB): the upper bound of the ML decoding obtained from Theorem 3. The integers in the legend denote the code lengths of QR codes.}
  \label{fig:47-71}
\end{figure}

 A comparison of the FER performance of HD, SD
and ML decoders for the (47, 24, 11) and the (71, 36, 11) QR codes
is made in Fig. \ref{fig:47-71}, and similar results can be drawn.
The (47, 24, 11) QR code significantly outperforms the (71, 36,
11) QR code when the HD decoding algorithms are utilized, In
contrast, the latter is approximately 0.3 dB better than the
former under the ML decoding. It is believable that the SD
decoding of the (47, 24, 11) QR code is also superior to that of
the (71, 36, 11) QR code even if lower bounds of SD decoding
rather than simulation results for two QR codes are given.

\begin{figure}[htbp]
\centerline{
\includegraphics[width=4.0in]{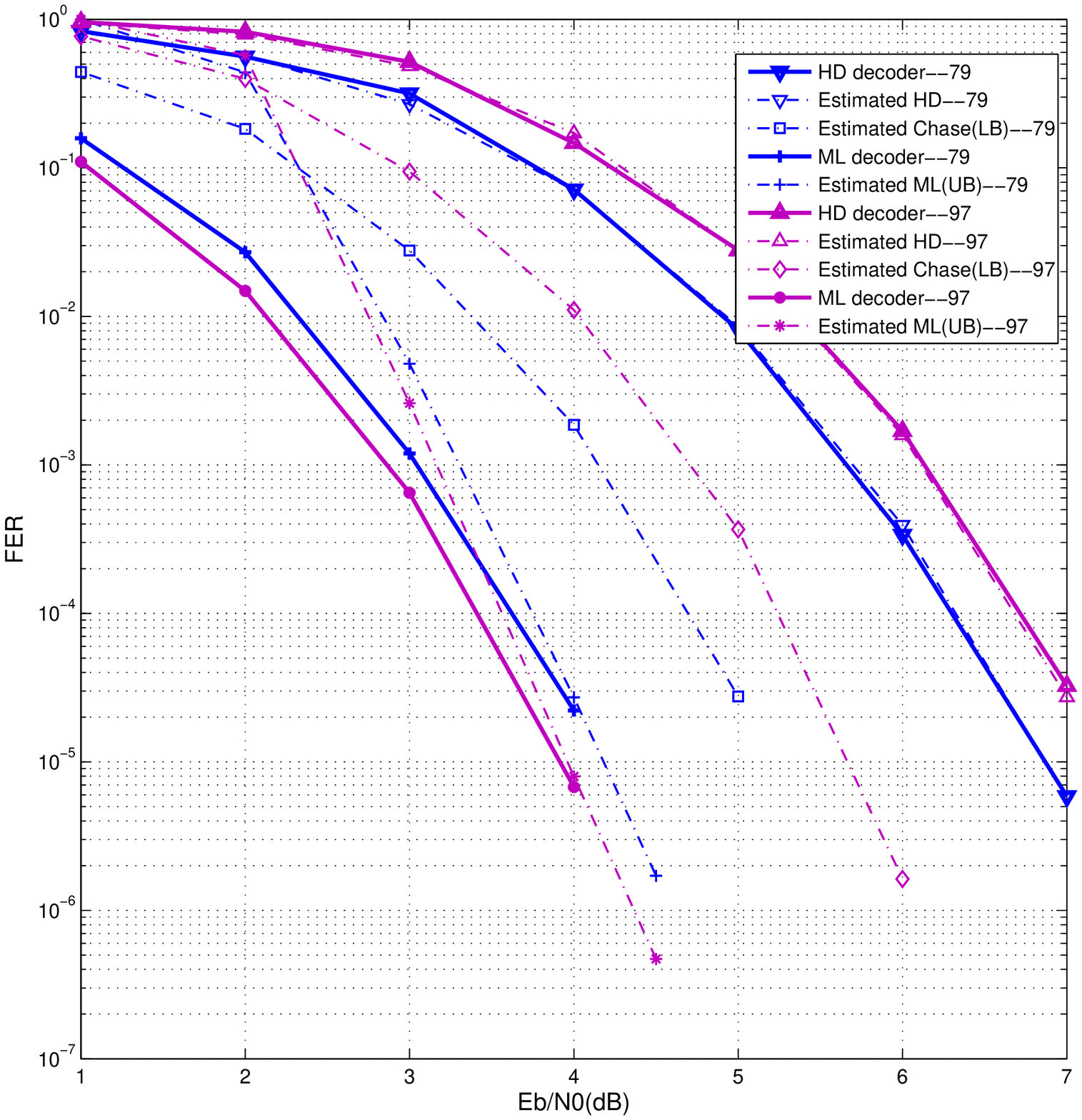}}
  \caption{FER performance comparison of the (79, 40, 15) and the (97, 49, 15) QR codes in AWGN channel. Estimated Chase (LB): the lower bound of the Chase decoding
  obtained from Algorithm 2; Estimated ML (UB): the upper bound of the ML decoding obtained from Theorem 3. The integers in the legend denote the code lengths of QR codes.}
  \label{fig:79-97}
\end{figure}

Performance comparison for the (79, 40, 15) and the (97, 49, 15)
QR codes are shown in Fig. \ref{fig:79-97}. As illustrated in Fig.
\ref{fig:79-97}, the HD decoding of the (79, 40, 15) QR code is
about 0.4 dB better than that of the (97, 49, 15) QR code at FER =
$1 \times 10^{-4}$. One can expect that the SD decoding of the
former provides a better performance even if only lower bounds of
performance rather than real performance are given. One also
observes that the SD decoding is more than 1.0 dB away from the ML
decoding for both codes, and $Theorem$ \ref{prop:PE_MLD} predicts
the ML decoding performance in high SNR regimes very accurately.

\begin{figure}[htbp]
\centerline{
\includegraphics[width=4.0in]{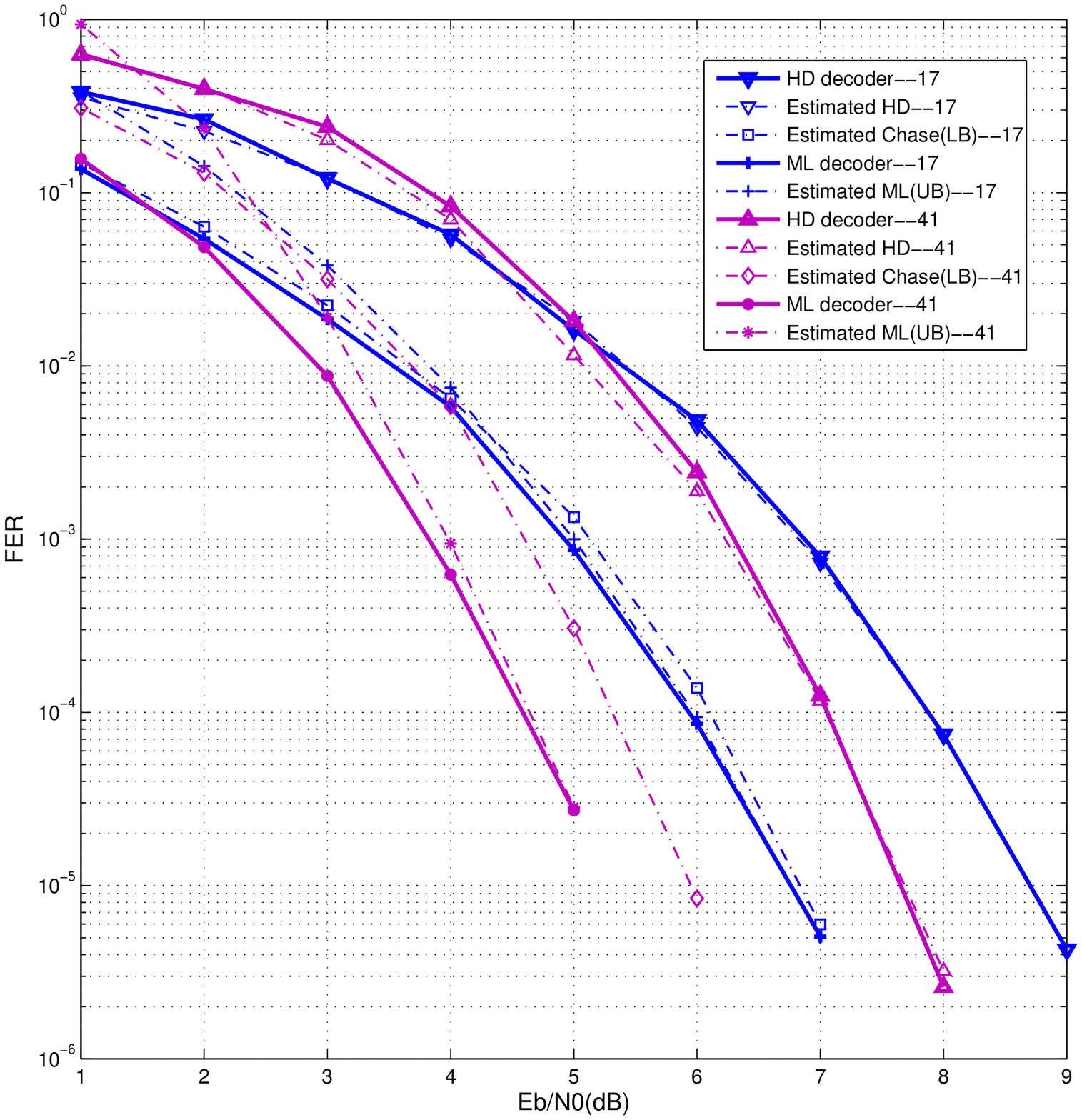}}
  \caption{FER performance comparison of the (17, 9, 5) and the (41, 21, 9) QR codes in AWGN channel. Estimated Chase (LB): the lower bound of the Chase decoding
  obtained from Algorithm 2; Estimated ML (UB): the upper bound of the ML decoding obtained from Theorem 3. The integers in the legend denote the code lengths of QR codes.}
  \label{fig:17-41}
\end{figure}
\begin{figure}[htbp]
\centerline{
\includegraphics[width=4.0in]{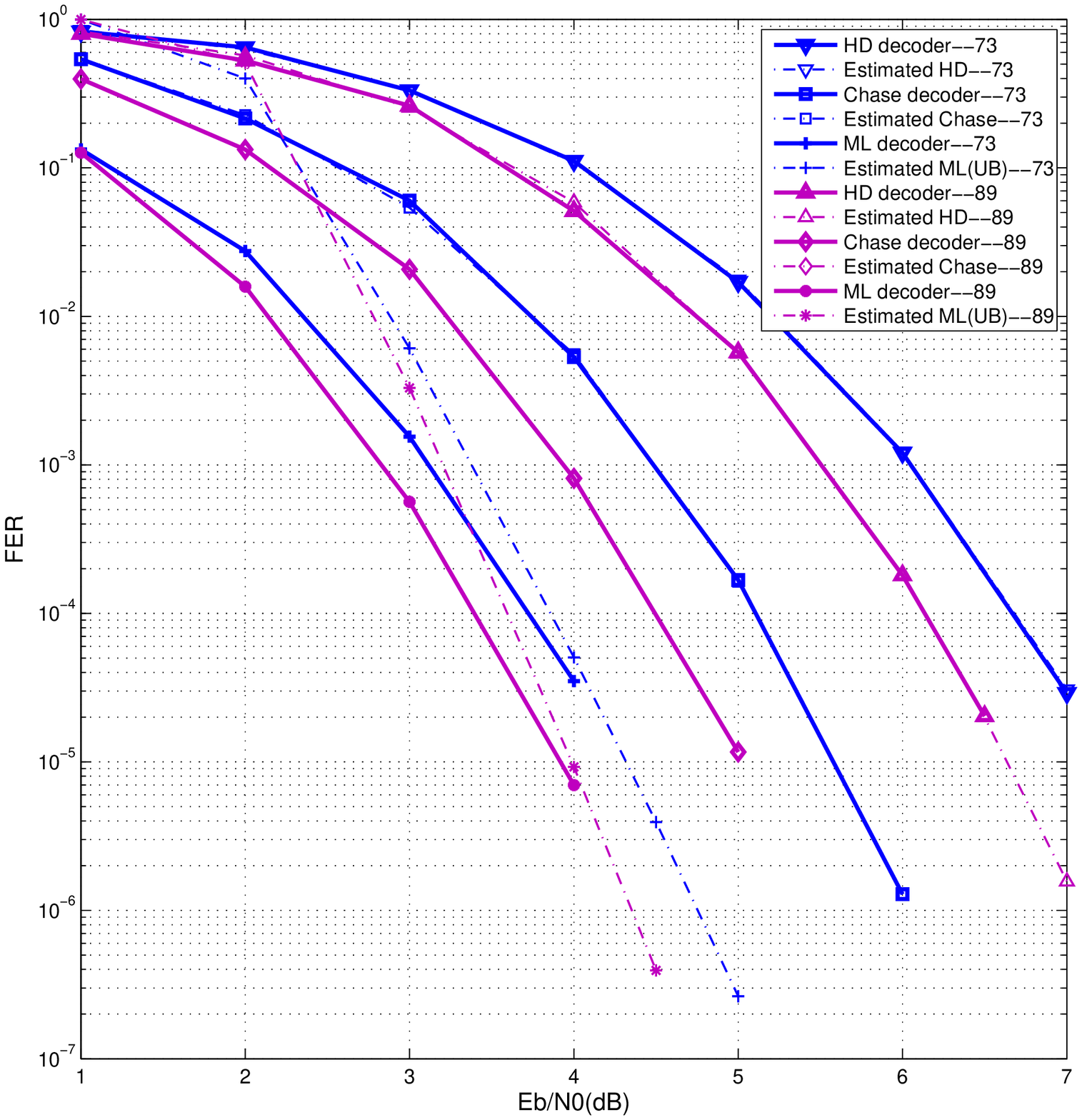}}
  \caption{FER performance comparison of the (73, 37, 13) and the (89, 45, 17) QR codes in AWGN channel. Estimated Chase (LB): the lower bound of the Chase decoding
  obtained from Algorithm 2; Estimated ML (UB): the upper bound of the ML decoding obtained from Theorem 3. The integers in the legend denote the code lengths of QR codes.}
  \label{fig:73-89}
\end{figure}

The performance of remainder four QR codes, i.e., QR codes with
lengths of 17, 41, 73, and 89, are shown in Fig. \ref{fig:17-41}
and Fig. \ref{fig:73-89}. It can be known that there only exists a
small gap between  the lower bound of performance of the Chase
decoder and  the ML decoder for the (17, 9, 5) QR code, whereas
such a gap widens to about 0.8 dB for the (41, 21, 9) QR code.
Also, one observes that the estimated method of the Chase decoder
has the same performance as the simulated method for the (73, 37,
13) and (89, 45, 17) QR codes. It is believable that the lower
bounds of the Chase decoding derived from the simple method
described in section III are very tight for the above considered
QR codes except the (23, 12, 7) QR code. Actually, the lower bound
for the (23, 12, 7) QR code is loose because it is a perfect code
and thus the probability of a received vector being decoded to a
valid but incorrect codeword increases when compared with other
imperfect codes. Moreover, the (89, 45, 17) QR code exhibits the
best HD and SD decoding performance since its minimum distance is
largest among tested 10 QR codes.

In a word, one observes from the above five figures that the SD
decoding is approximately 1.5 dB superior to the HD decoding for
each QR code of length within 100. Furthermore, one also observes
that the gap between the SD decoder and the ML decoder gradually
widens as the code length increases. $Theorem$ \ref{prop:PE_MLD}
is reliable for calculating the ML decoding performance in the
high SNR regimes for tested 10 QR codes, and thus can be used to
predict the performance of more longer QR codes under the ML
decoding without simulations. Additionally, the HD decoding
performance can also be estimated rapidly according to $Theorem$
\ref{prop:HD}, and the estimated curves matches the simulation
results very well in the whole simulated SNR regions. As a result,
Theorems \ref{prop:HD} and
 \ref{prop:PE_MLD} provide a simple and efficient method to
predict and analyze the performance of QR codes with lengths
beyond 100.

\begin{figure}[htbp]
\centerline{
\includegraphics[width=6.5in]{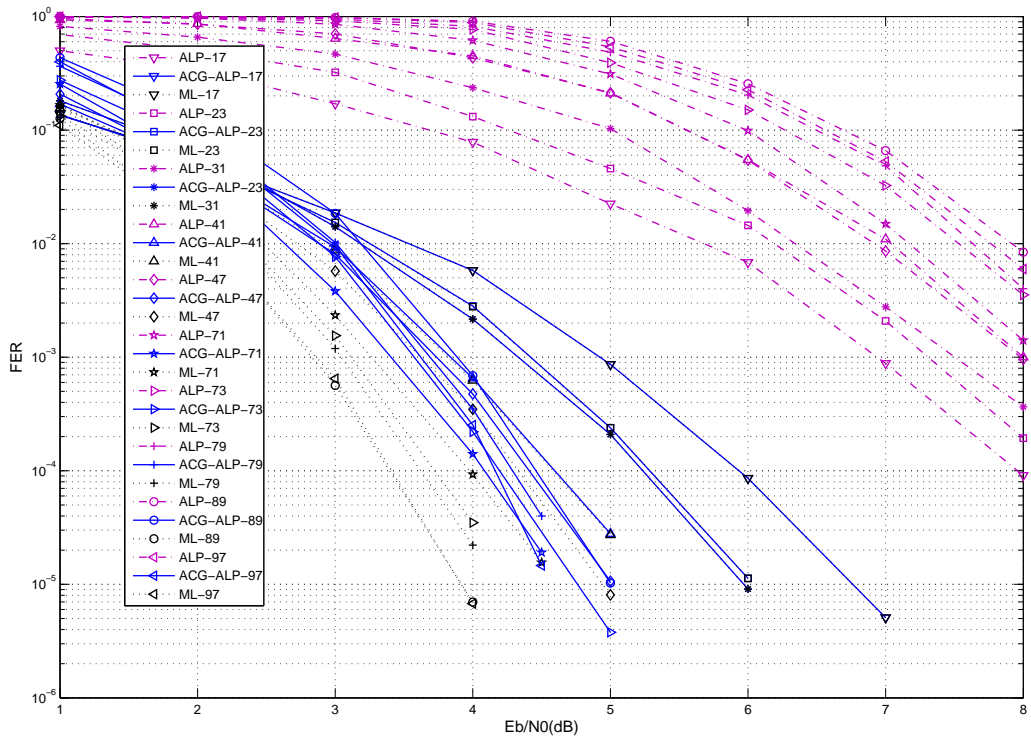}}
  \caption{FER performance of 10 QR codes mentioned previously, where ALP-$n$ (dot-dashed lines), ACG-ALP-$n$ (solid lines), and ML-$n$ (dotted lines) represent the
  ALP decoding, the ACG-ALP decoding and the ML decoding, respectively, for the QR code of length $n$.}
  \label{fig:LP_decoding}
\end{figure}

We also investigate the performance of the aforementioned 11 QR
codes except for the (7,4,3) QR code when using different LP-based
decoding algorithms. Herein, the ALP decoder \cite{Taghavi}
provides the same error-correcting performance as the standard LP
decoder proposed by Feldman et al.\cite{Feldman} with reduced
computational complexity, while the ACG-ALP decoder has best
performance among the known LP-based decoders \cite{Zhang,
Wang2013}. Upon inspection of Fig. \ref{fig:LP_decoding}, the ALP
decoder has very poor performance for each QR code especially for
long codes. It seems to be confused that the performance of the
ALP decoder becomes worse when the code length increases. We guess
it is because the ALP decoder converges to a pseudo-codeword more
likely on a larger polytope and thus deteriorates the performance.
Whereas, the ACG-ALP decoder always performs very well by using
redundant parity-check constraints. For the (17, 9, 5), (23, 12,
7), (31, 16, 7), and (41, 21, 9) four QR codes, the ACG-ALP
decoder performs as well as the ML decoder. But there more or less
exists a gap between the ACG-ALP and ML decoders for anyone of the
remainder six codes, and  the
 ACG-ALP decoder is farther away from the ML decoder
when the code length increases. Among the tested 10 QR codes, the
(71, 36, 11) rather than the (89, 45, 17) QR code is the best one
under the ACG-ALP decoding although the latter has the largest
minimum distance. In fact, the (73, 37, 13) also outperforms the
(89, 45, 17) QR code when decoding by the ACG-ALP algorithm. Such
a phenomenon was partly explained by pseudo-codeword frequency
spectrum analysis in \cite{Li2014}, where the pseudo-codeword
frequency spectrum of the (73, 37, 13) and the (89, 45, 19) QR
codes are compared. However, further research needs to be
investigated in order to reveal the internal reasons.

It is well-known that both the ACG-ALP and the algebraic SD
decoders are based on soft information. Consider the simulated 10
QR codes, A conclusion can be drawn by comparing the above
simulation
 figures that the ACG-ALP decoder performs at least as well as the
 SD decoder. For some QR codes such as the (73, 37, 13) QR code, the former
significantly outperforms the latter. However, it is difficult to
compare the detailed arithmetic operations between the ACG-ALP
decoder and the algebraic SD decoder. The computational complexity
of the ACG-ALP decoder is mainly decided by two factors: the
number of iterations (i.e., the number of LP problems) of decoding
a codeword, and the complexity of the LP problem in each
iteration. In our simulations, the simplex algorithm is utilized
as the LP solver. Such an algorithm often converges rather rapidly
although it has an exponential worst-case complexity. The simplex
method needs $2n \sim 3n$ pivot steps typically \cite{Szabo} and
requires O($n^{3}$) operations in each pivot step, where $n$
denotes the number of primal variables, i.e., the code length.
Thus the ACG-ALP decoder requires O($c\cdot n^{4}$) operations to
decode a codeword, where $c$ is the average number of iterations.
As described in Section III, when $v$ errors occur and Gaussian
elimination is used, the HD decoding has a complexity of
O($q\cdot(v+1)^{3} + 3t + v\rho $), where $q, v, \rho$ indicate
the number of elements in the finite field on which a QR code is
defined, the number of occurred errors and the code length,
respectively. In low SNR region, the weights of most of error
patterns are larger than $t$, which contradicts the condition (13)
in \cite{Wang2013}. Therefore, the SD decoding cannot be
terminated until it exhausts all the $2^{t}$ possible error
patterns, i.e., the HD decoding is repeated recursively for
$2^{t}$ times. The complexity of the SD decoder requires
O($(q\cdot (t+1)^{3} + 3t + t\rho)\cdot 2^{t}$). However, in the
high SNR regime, fewer errors occur in the received vector and the
algebraic SD decoding algorithm can often be terminated quickly by
the sufficient optimality condition in \cite{Shu Lin, Wang2013}
because of obtaining an ML codeword. Thus, its computational
complexity can be reduced to O($((v+1)! + 3t + v\rho )\cdot k$),
where $k \ll 2^{t}$. It should be noted that we use the real field
to compute the ACG-ALP decoder and the finite field to compute the
algebraic SD decoder. In the finite field, the arithmetic
operations refer to look-up tables and modular operations
additionally.

Recall that the  complexity comparison mentioned previously is
  not based on hard-and-fast complexity estimates since it seems
  to be very difficult to measure the algorithm
  complexity of the ACG-ALP decoder in terms of Big-Oh estimates.
In our previous simulations \cite{Wang2013, Li2014}, the ACG-ALP
decoder runs faster than the algebraic SD decoder in the low SNR
regions, but in the high SNR regimes, the former is slower than
the latter. Hence, we can simply draw a conclusion that the total
computational complexity of both decoders are comparable.
Actually, the speed of the ACG-ALP decoder can be further improved
once other powerful LP solver such as interior-point solver is
utilized.

\section{Conclusion}
  In this paper, three algorithms consisting of HD, SD
  and LP decoders for decoding QR codes of lengths within 100 are
  investigated. For short QR codes, the error-locator polynomials
  can be found by solving Newton identities. In contrast, the inverse-free
  BM algorithm is more efficient for long QR codes since
  determining the coefficients of the error-locator polynomial by
  solving Newton identities is very difficult in this case. The SD decoding can be
  achieved by combining the HD decoding together with the Chase-II algorithm. The LP
  decoding, however, can be easily conducted by representing a QR code with
  a parity-check matrix. Using an
  excellent LP-based decoder, namely ACG-ALP decoder, an efficient
  algorithm is developed to obtain the performance of the ML decoding
  for QR codes with lengths less than or equal to 100.

  Under the all-zero codeword assumption, an equation is derived
  to estimate the FER performance of the HD decoding. It is in agreement with the simulation
  results in the whole simulated SNR regimes for all the
  tested QR codes.  A simple simulation-based method is proposed to
  estimate the SD  decoding performance, which provides a lower bound actually by
  assuming that there exists no undetected error. Also, a
  mathematical formula is proposed to calculate the ML decoding
  performance of QR codes. Such a formula coincides with
  simulation results for moderate to high SNR regimes and can
  usually employ any cyclic code once its weight numerator is
  given.

In this paper, two important theorems are proposed to explore some
internal properties of QR codes. In other words, given two QR
codes with the same minimum distances, the shorter code performs
better than the longer one under the HD decoding, whereas the
latter is more excellent when the ML decoding is used. It is
expected that such two theorems together with the above methods of
estimating performance of different decoding algorithms can be
efficiently utilized to analyze and predict the decoding
performance of QR codes whose lengths are beyond 100.

\section*{Acknowledgment}

The authors would like to thank Dr. H.-C. Chang for helpful
discussions on the Newton identities.

\ifCLASSOPTIONcaptionsoff
  \newpage
\fi



%

\newpage

\end{document}